\documentclass[conference,a4paper]{IEEEtran} 
\usepackage{graphicx}
\usepackage{cite}
\usepackage{verbatim}
\usepackage{cite}
\usepackage{url}
\usepackage{graphics}
\usepackage{graphicx}
\usepackage{hyperref}
\usepackage{color}

\begin{document}
	
\title{SDN based Control and Management of WLANs in the 3GPP 5G Network}	
\author{
\IEEEauthorblockN{ Pranav Jha, Abhay Karandikar}
\IEEEauthorblockA{Department of Electrical Engineering, Indian Institute of Technology Bombay \\
Email: {$\lbrace$pranavjha, karandi$\rbrace$}@ee.iitb.ac.in}
}
		
	
\maketitle
	
\begin{abstract}
The exponential growth in mobile broadband usage~\cite{cerwall2017ericsson} has catalyzed the need for high data rate communication systems. In this regard, activities for standardizing the next-generation mobile broadband system, known as the Fifth Generation (5G) system are underway. The 5G system also enables the integration of Institute of Electrical and Electronic Engineers (IEEE) Wireless Local Area Networks (WLANs) for providing cost-effective broadband connectivity. It is therefore imperative to find solutions for control and management of WLANs, while providing seamless inter-working capabilities with the cellular network. In this paper, we propose a novel Software Defined Networking (SDN) based architecture for efficient control and management of IEEE WLANs while providing a mechanism for smooth integration of WLANs within the 5G system.
%
%
\end{abstract}

\section{Introduction}

Broadband mobile data usage has experienced a tremendous growth in the past decade, giving rise to multiple standardization initiatives towards the design and development of higher capacity networks known as the Fifth Generation (5G) networks. The 5G network which is being developed by the Third Generation Partnership Project (3GPP), is one such key standard.

\par In addition to the development of new wireless communication technologies, other measures such as the deployment of IEEE 802.11 based Wireless Local Area Networks (WLANs) alongside the cellular network are also being undertaken. WLANs are particularly suitable for low mobility access scenarios, e.g., to support stationary or pedestrian users and are a cost-effective mobile broadband access technology. Therefore, in areas where cellular coverage may be poor or data demand is very high, operators are supplementing their cellular networks with IEEE 802.11 based WLANs. At present, WLAN equipment vendors provide proprietary mechanisms for Access Point (AP) management. This leads to interoperability issues across APs from different vendors. An attempt to address these issues was  made by the Internet Engineering Task Force (IETF) by standardizing a protocol known as Control And Provisioning of Wireless Access Points (CAPWAP)~\cite{stanley2009control}. CAPWAP provides a centralized control and management mechanism for WLAN APs and helps in achieving interoperability across APs from different vendors. However, this protocol does not provide a well-defined programmable interface over the APs and therefore not in agreement with the SDN principles.

\par Along with these developments in wireless communication technology, significant progress is also being made towards the evolution of newer architectural and design principles for the next-generation communication systems. Software Defined Networking (SDN) is one such emerging architectural paradigm, which is transforming the approach for control and management of a communication network. It promises to provide significant advantages over conventional networks in terms of flexibility, programmability, vendor interoperability, optimized resource utilization, and granular control over the deployment of policies/services in the network.

\par SDN enables the separation of control and forwarding (data) plane functions in a network through a standardized programmable interface~\cite{mckeown2008openflow}. This is made possible by transposing the control plane functionality from the network devices such as routers, switches etc., to SDN Controllers in a logically centralized fashion. These modified network devices form a part of the forwarding plane in the network and are responsible for forwarding data. The data forwarding decisions in the devices are taken on the basis of the rules provided by the SDN Controller. In this paper, we propose an SDN based architecture for control and management of IEEE 802.11 WLANs. The proposed architecture brings the advantages of SDN to IEEE 802.11 based Radio Access Network (RAN).

\par Due to increasing WLAN deployments alongside the 3GPP cellular networks, there is an increasing need for unified control and management of 3GPP cellular networks and non-3GPP access networks (IEEE 802.11 based WLANs). 3GPP has made continued efforts for efficient inter-working of WLANs with 3GPP cellular networks. Multiple enhancements to the 3GPP Long Term Evolution (LTE) standard, across different 3GPP releases, have been made in order to make it inter-working with WLANs~\cite{rajavelsamy2015review}. The proposed 3GPP standard for the 5G System~\cite{5gspec}, i.e., the Release 15 of the 3GPP standard, aims to enable inter-working between WLAN based access network and the 5G Core (5GC) network through a logically independent inter-working function known as the Non 3GPP Inter-Working Function (N3IWF). This function is still under standardization in 3GPP. In this paper, we also present an SDN based technique for implementing the inter-working functions for seamless integration of WLAN with the 5GC.

\par The rest of the paper is organized as follows. Section II discusses some of the available literature in this area. Section III provides details regarding the 3GPP defined 5G architecture. The proposed SDN based architecture is discussed in Section IV. The subsequent section highlights the advantages of the proposed architecture. Section VI concludes the paper and indicates the directions for future work.

\section{Related Work}
In this section, we present a brief summary of the existing literature in this area. Some of them~\cite{schulz2014programmatic,jang2017rflow+} define mechanisms for SDN based control and management of WLANs. Authors in \cite{schulz2014programmatic} propose a solution based on Lightweight Virtual APs (LVAPs) which can be created and controlled by a centralized SDN Controller. The solution combines two south bound protocols viz., OpenFlow for traffic steering and Odin for managing the wireless radio. Another work \cite{jang2017rflow+} provides a scalable SDN-based monitoring and management framework for WLAN. 
\par The second set of articles explore the integration of WLANs with the Fourth Generation (4G) 3GPP LTE systems.  Authors in \cite{wang2016sdn} propose an SDN based converged architecture for LTE and WLAN. This architecture provides  seamless mobility at the service level between LTE and WLAN using virtualization. This solution involves the use of a virtual middlebox in the UE for choosing the network interface to/from which data is to be steered and route the control messages.

\section{3GPP Defined 5G Architecture}

\begin{figure}
	\centering
	\includegraphics[width=0.45\textwidth]{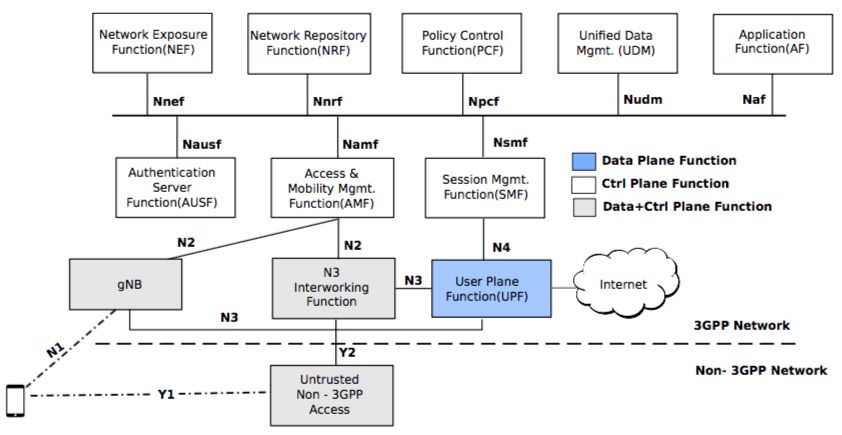}
	\caption{3GPP Defined 5G architecture}
	\label{fig:temp_3gpp_5G}
\end{figure}

This section provides details on the 3GPP-defined 5G system architecture. This architecture, illustrated in Fig. \ref{fig:temp_3gpp_5G} can be broadly divided into two parts viz., the 5GC and the access network known as the Next Generation Radio Access Network (NG-RAN). The access network supports a multitude of access technologies, both, 3GPP and non-3GPP. The 5GC is common for all of these diverse access networks. The 5G system is constituted as a set of Network Functions (NFs)~\cite{5gspec}. NFs are entities with well-defined functionality and interfaces. The NFs in the 5GC follow the SDN paradigm and are identifiable either as control or data plane functions. These NFs interact with each other over standardized interfaces. However, the 3GPP access network nodes belonging to the NG-RAN, e.g., gNB and the next generation eNodeB (ng-eNB) incorporate both the control and the data plane functionality without a standardized programmable interface between the two planes. The NG-RAN may comprise of NR based radio access nodes called the gNBs or LTE based next-generation radio access nodes known as the ng-eNBs. A gNB provides NR data and control plane protocol interface towards the UE and an ng-eNB, provides the Evolved Universal Terrestrial Radio Access (E-UTRA) data and control plane protocol interface. For the sake of brevity, this paper uses only the gNB as the reference.

\begin{figure}
	\centering
	\includegraphics[width=0.9\linewidth]{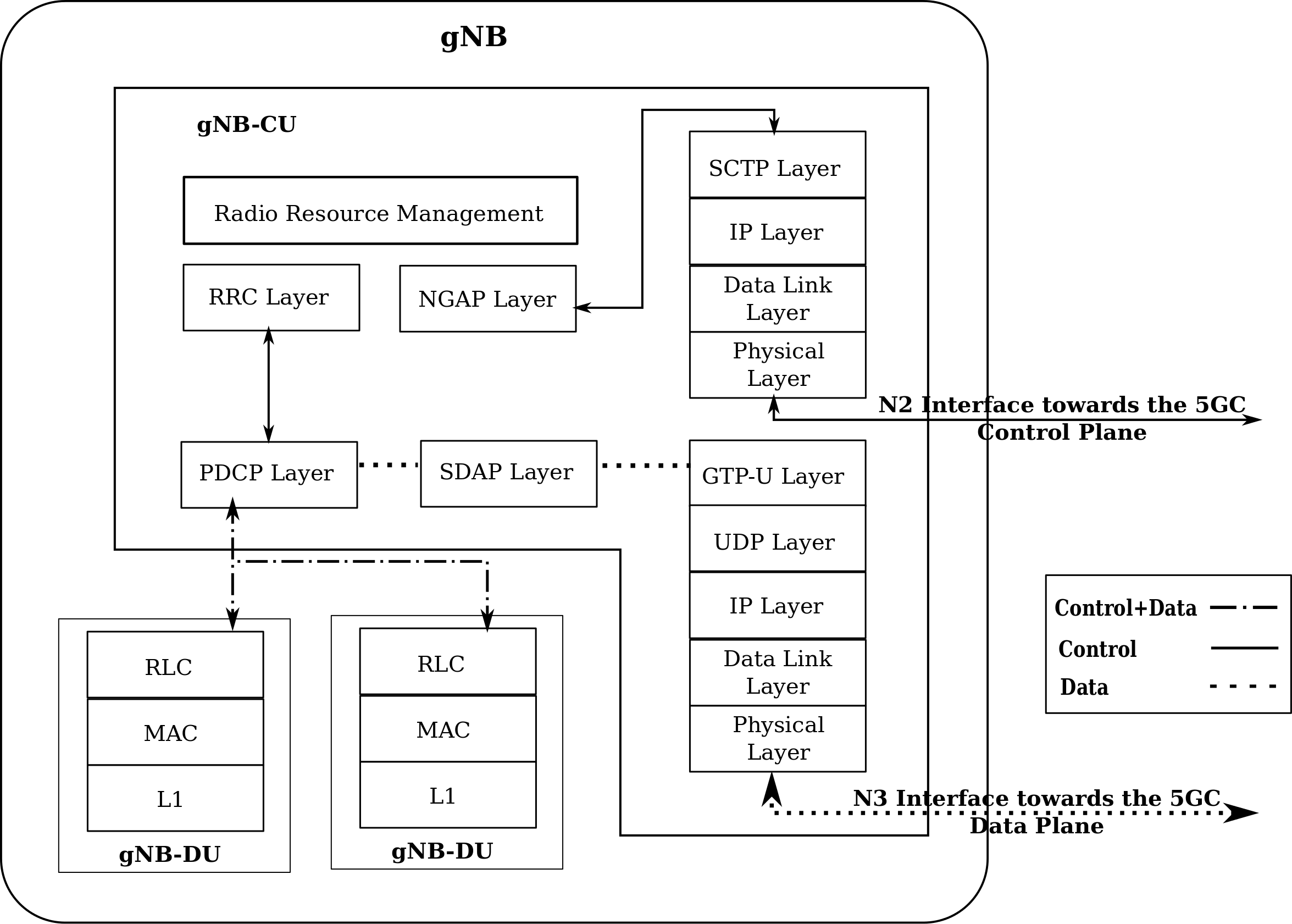}
	\caption{Protocol Stack of the 3GPP gNB}
	\label{fig:3gppgNB}
\end{figure}

\par The gNB as illustrated in Fig. \ref{fig:3gppgNB} consists of two types of units, a gNB Central Unit (gNB-CU) and one or more gNB Distributed Units (gNB-DUs). A gNB-CU is a logical node, which controls the operation of one or more gNB-DUs. It hosts the control plane protocol as well as a few data plane protocols on the radio interface. It also interfaces with the 5GC for both, data plane (over N3 interface) and control plane over the N2 interface. The gNB-CU also comprises of the Radio Resource Management (RRM) functionality, e.g., admission control, connected mode mobility control, and radio bearer control functionality. The gNB-DU is a logical node hosting the Radio Link Control (RLC), Medium Access Control (MAC) and Physical Layers of the gNB. It is partly controlled by a gNB-CU. Every gNB-DU supports one or more cells. Although the UE access and session management functionality is handled by the 5GC network in a unified manner, RRM functions are likely to be handled at individual access network entities, i.e., at a gNB-CU, possibly leading to sub-optimal decisions. The N3IWF connects the 5GC to the non-3GPP access networks. 

\section{Proposed Architecture}
\begin{figure}
	\centering
	\includegraphics[width=\linewidth]{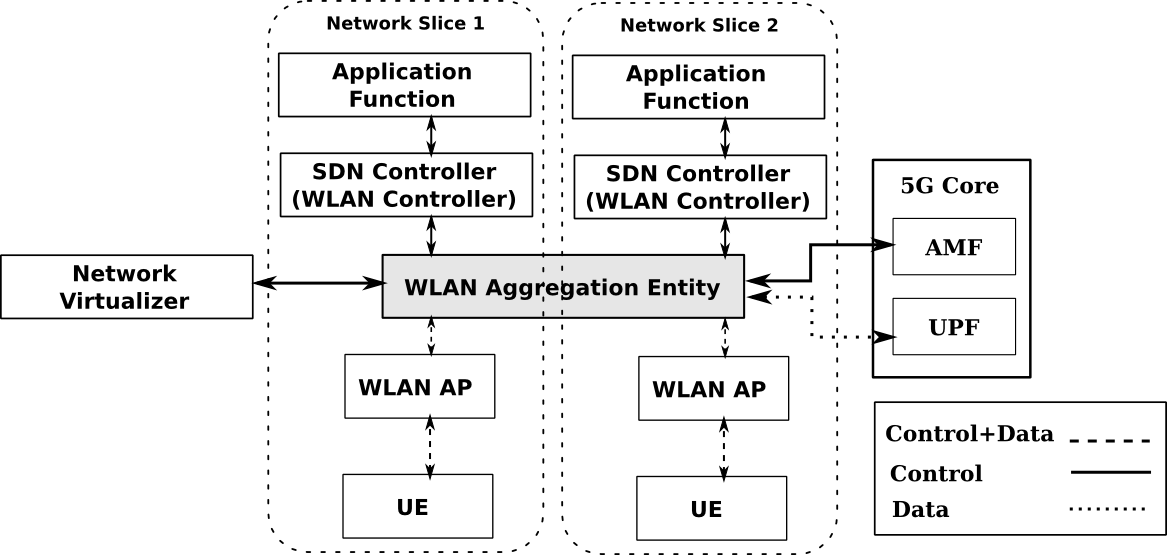}
	\caption{System Architecture of the Proposed SDN based WLAN Controller}
	\label{fig:wlan5garch}
\end{figure}

In this section, we propose an SDN based architecture for control and management of IEEE 802.11 based WLANs. The proposed architecture has been depicted in Fig.~\ref{fig:wlan5garch}. This architecture separates the control plane functions in WLANs from the forwarding (data) plane functions. It also enables the seamless integration of WLANs with the 5GC with the help of a WLAN Aggregation Entity (WAE).
\par The proposed WLAN architecture consists of the following logical entities: WLAN APs, WAE, the SDN based WLAN Controller (also called the RAN Controller), Application Function (AF) and the Network Virtualizer (NV). The RAN Controller together with the AF implements the control plane functionality in the network. NV is also a control plane entity. However, it is primarily responsible for virtualizing the physical network and facilitating the creation of multiple logical networks over a physical network. The WAE and the WLAN APs constitute the data plane in the network.
\par The functionality of the various component functions in the control and data plane are described in detail below:
\begin{enumerate}
\item \textbf{Control Plane Functions}
\begin{itemize}
\item NV: The NV function creates multiple logical networks, also called known as `network slices' over a single physical RAN. Each of these logical networks or network slices may comprise of individual data and/or control plane entities/functions as shown in Fig.~\ref{fig:wlan5garch}.
\item AF: AF is responsible for the implementation of the RAN resource management algorithms, such as load balancing, admission control and mobility management, spectrum management etc.. It is integrated with the  RAN Controller through a standardized Interface.
\item RAN Controller: The RAN Controller is responsible for the following functionalities in the control plane. It supplies configuration to the data plane entities and also performs the mapping from the UE to the AP. The RAN Controller also manages the RAN specific control plane communication for the UEs. It provides the Quality of Service (QoS) attributes to the RAN data plane entities for an individual  traffic flow. This can be achieved with the help of the following sub functions viz., the RAN Management Function (RMF), Flow Control Function (FCF), RAT Control Function (RCF) and the Data Plane Interface Function (DPIF). The RMF provides the relevant entity specific configuration parameters to the data plane entities. The FCF  deals with an abstract view of the underlying network in terms of UE specific traffic flows and is responsible for setting up these flows with requisite Quality of Service (QoS) attributes through the RAN data plane entities. The flow specific QoS attributes are received by the RAN Controller from the 5GC. The RCF exchanges RAN specific control messages with the underlying data plane entities for every UE and manages the UE to AP mappings. This mapping is established under the control of algorithms running as part of the AF. The DPIF comprises of the NETCONF and OpenFlow protocols and acts as the RAN Controller interface towards the data plane i.e., the WAE.
\end{itemize}
\item \textbf{Data Plane Functions}
\begin{itemize}
\item WAE: The WAE is a data plane aggregation function built over the WLAN APs. It presents a unified management and control interface of the underlying RAN to the control plane. It receives relevant control and management information from the RAN Controller and provides them to the APs and UEs. A NETCONF and OpenFlow protocol based standardized interface is used for the exchange of information between the RAN Controller and the WAE. The WAE also facilitates integration of WLANs with the 3GPP 5GC. It relays data and UE specific control (Non-Access Stratum (NAS) signaling) messages exchanged between UEs and the 5GC. The relayed control messages include information for the authentication of UEs. It also establishes an Internet Protocol Security (IPSec) tunnel with the UEs for secure communication between the two. This entity also segregates the data path from the control path. Control signals (NAS signaling messages) are exchanged with the 5GC network over the N2 interface, whereas the user data is exchanged over the N3 interface with the 5GC. It also applies the required QoS attributes to the individual data flows in the RAN. This entity also acts as the mobility anchor for the UEs, i.e., the WAE for a particular UE does not change when the UE moves across APs.
\item WLAN AP: The WLAN APs are connected to the UEs on one side and the WAE on the other side and is responsible for relaying data/UE specific control signaling exchanged between UEs and the 5GC via the WAE. The APs communicate with the UEs over the IEEE 802.11 based radio interface. The APs also exchange RAN specific control information with RAN controller via the WAE. The RAN Controller shares RAN specific control information with APs to control the UE-to-AP mappings and also broadcast information by APs in the basic service area.
\end{itemize}
\end{enumerate}
\section{Advantages of the Proposed Architecture}
The proposed architecture provides multiple advantages, some of which are described below:
\subsection{Enables Slicing of the Network}
Due to the increasing diversity of services, with each service having its own performance requirements, it is difficult for operators to deliver these services over same physical infrastructure. Network slicing resolves this problem by dividing the physical infrastructure into smaller logical networks known as slices. The proposed architecture enables network slicing support for WLAN networks at the RAN. The RAN Controller facilitates the complete life cycle viz., the Create, Read, Update, Delete (CRUD) operations for network slice templates to support diverse applications. Network slice templates are data filters that help in mapping a particular service to a given slice instance. The RAN Controller configures and manages the resources in WLAN APs and the DPIF, based on the slice template. It provides efficient end-to-end service orchestration of RAN deployments, services for network scalability etc.. Slicing also enables multi-tenancy support and flexibility in the network.

\section{Conclusion}
In this paper we have proposed an SDN based architecture for control and management of WLAN. This architecture also provides a novel mechanism for enabling integration with the proposed 3GPP-defined 5G network. We have also compared other existing solutions and present possible use cases for the proposed architecture.

\section{Acknowledgments}
This research work has been funded through a grant on SDN and 5G from Department of Electronics and Information Technology (DeITY), Government of India. We thank Abhishek Dandekar, Ashish Sharma, Rohan Kharade and Akshatha Nayak M. for their contributions to this paper.

\bibliographystyle{ieeetr}
\bibliography{wlansdn}

\end{document}